\documentclass[amsmath,amssymb]{revtex4}
 
\usepackage{graphicx}
\usepackage{dcolumn}
\usepackage{bm}
\begin{document}
\title{Detection of invisible and crucial events: 
from seismic fluctuations to the war against terrorism }
\author{Paolo Allegrini$^1$, Leone Fronzoni$^{2,3}$,
Paolo Grigolini$^{2,4,5}$, Vito Latora$^{6}$, Mirko S. Mega$^{2}$,
Luigi Palatella$^{2}$, Andrea Rapisarda$^{6}$ and Sergio Vinciguerra$^{6,7}$}

\affiliation{$^1$Istituto di Linguistica Computazionale del Consiglio Nazionale delle
Ricerche, Area della Ricerca di Pisa, Via Moruzzi 1, San Cataldo,
56010, Ghezzano-Pisa, Italy}
\affiliation{$^{2}$Dipartimento di Fisica dell'Universit\`a di Pisa and
   INFM, via Buonarroti 2, 56127 Pisa, Italy}
\affiliation{$^{3}$Centro Interdipartimentale per lo Studio dei
  Sistemi Complessi, via S. Maria 28, 56126 Pisa, Italy}
\affiliation{$^{4}$Center for Nonlinear Science, University of North Texas,
   P.O. Box 311427, Denton, Texas 76203-1427 }
\affiliation{$^5$Istituto dei Processi Chimico Fisici del CNR
Area della Ricerca di Pisa, Via G. Moruzzi 1,
56124 Pisa, Italy}
\affiliation{$^{6}$ Dipartimento di Fisica e Astronomia, Universit\`a di 
   Catania, and INFN, Via S. Sofia 64, 95123 Catania, Italy}
\affiliation{$^7$ Osservatorio Vesuviano - INGV, Via Diocleziano 328, 
80124 Napoli, Italy }
\begin{abstract}
 We argue that the recent discovery of the non-Poissonian statistics
 of the seismic main-shocks is a special case of a more general
 approach to the detection of the distribution of the time increments
 between one crucial but invisible event and the next. We make the
 conjecture that the proposed approach can be applied to the analysis
 of terrorist network with significant benefits for the Inteligence
 Community.
\end{abstract}
\maketitle

\section{Introduction}

The main aim of this paper is to prove the efficiency of a new method
for the detection of crucial events that might have useful
applications to the war against terrorism. This has to do with the
search for rare but significant events, a theme of research that has
been made of extreme importance by the tragedy of September 11. This
method is applied here to defining the statistics of seismic
main-shocks, as done in an earlier publication
\cite{earlier}. However, the emphasis here is more on the conceptual
issues behind the interesting results obtained in Ref. \cite{earlier}
than on their geophysical significance.  In fact, the discussion of
these conceptual issues aims at supporting the conjecture that the
method has a wider range of validity. We shall help the reader to
understand this general discussion with a dynamic model, originally
proposed in Ref. \cite{memorybeyondmemory}.  We point out that this
model was proposed for purposes different from geophysical
applications. However, it is a case where the crucial events to detect
are under our control, thereby making it possible for us to check the
accuracy of the method of detection of invisible and crucial events
that we propose here for a more general purpose, including the war
against terrorism.  Furthermore, for this model an analytical
treatment has been recently found \cite{memory2}, supporting the
claims that we make in this paper for the accuracy of the method of
detection.  For the reader's convenience, the results on the seismic
fluctuations are suitably reviewed, and discussed in the light of the
more general perspective of this paper.  We also review the model for
seismic fluctuations, proposed in the earlier work of
Ref. \cite{earlier}. This model shares with the model of
Ref.\cite{memorybeyondmemory} the property that the crucial events are
imbedded in a sea of secondary events, but it allows us to reveal with
accuracy the statistics of the crucial events for different
mathematical reasons.

\section{Crucial events, memory and predictability}

The analysis shown in action on the seismic fluctuations, should serve
the more general purpose of detecting the statistical properties of
\emph{crucial} events that are \emph{invisible}. By crucial events we
mean events causing other events, which would be predictable if the
time occurrence of their causes were known. By invisible and crucial
events we mean crucial events embedded in a sea of many other events,
either caused by the invisible events or by environmental
fluctuations. These secondary events play a camouflage action that
makes it difficult to detect the crucial events with accuracy. We
discuss first the property that the crucial events must have,
regardless of whether they are visible or not.  We want also to
address the delicate issue of the extent to which the crucial events
are predictable and the extent to which they are random, this being a
problem that has caused much confusion in the past.  Let us consider
the following dynamic model.  A particle moves in the interval $I
\equiv [0,1]$. Its trajectory $x(t)$ is governed by the equation:
\begin{equation}
\label{dynamics}
\frac{dx}{d t} = \alpha x ^{z}.
\end{equation}
The parameter $\alpha$ is a positive number, which has to be kept much
smaller than $1$, if the integration time step is $1$. This
fundamental equation serves the purpose of generating non-Poisson
statistics. In fact, as we shall see, $z=1$ generates Poisson
statistics, while the wide dominion of non-Poisson statistics is given
by $ z > 1$.  When the particle reaches the border $x = 1$ it is
injected back to a new initial condition $x (0) > 0$, with uniform
probability. Now, let us imagine that the times of sojourn within the
interval are recorded via direct observation. For tutorial purposes,
we assume these random events to be visible. We shall get the time
series
\begin{equation}
\label{timeseries}
\{\tau\} \equiv  \tau(1), \tau(2),  ...
\end{equation}
Let us consider the initial condition $x(0)$. The time spent by the
particle, moving from this initial condition, within the interval $I$,
before reaching the border is
\begin{equation}
\label{timeversusinitial}
\tau = \frac{1}{\alpha} \left [ \frac{1}{1-z} - \frac{x(0)^{1-z}}{1-z} \right ],
\end{equation} 
as predicted by the time integration of Eq. (\ref{dynamics}). The
connection between the waiting-time distribution and the injection
process into the initial condition is given by
\begin{equation}
\label{uniform}
\psi(\tau)d\tau= p(x(0))dx(0).
\end{equation}
In the case of a uniform injection, $p(x(0)) = 1$, inserting  
Eq. (\ref{timeversusinitial}) into Eq. (\ref{uniform}) yields
after some algebra
\begin{equation}
\label{waitingtimedistribution}
\psi(\tau) = (\mu -1) \frac{T^{\mu-1}}{(T + \tau)^{\mu}},
\end{equation}
where
\begin{equation}
\mu \equiv \frac{z}{z-1}
\end{equation}
and
\begin{equation}
T = \frac{1}{\alpha(z-1)}. 
\end{equation}
It is worth devoting some comments to this result. Let us imagine that
we convert the time serie $\{\tau\}$ into a diffusion process with the
same prescription as that adopted in the earlier work on the seismic
fluctations in Southern California. This means that the random walkers
makes a jump ahead by a given quantity, equal to $1$, for instance, at
the times $t(1) = \tau_{1}$, $t(2) = \tau_{1} + \tau_{2}$, $t(3) =
\tau_{1} + \tau_{2} + \tau_{3}$, and so on. An ensemble of random
walkers obeying the same prescription (see Ref. \cite{giacomo}, for
details on how to derive this ensemble from the single sequence
$\{\tau\}$) undergoes a diffusion process that in the specific case $2
<\mu < 3$ yields a diffusion process. The probability distribution
function, $p(x,t)$, for this process, in the time asymptotic limit is
expected to obey the scaling condition
\begin{equation}
\label{scaling0}
p(x,t) = \frac{1} {t^{\delta}} F(\frac{x - wt}{t^{\delta}}),
\end{equation}
where $w$ is the mean velocity produced by the walking rule adopted
and $\delta$ is the scaling index given by
\begin{equation}
\label{scaling}
\delta = \frac{1}{\mu -1} .
\end{equation}
$F(y)$ is an asymmetric function of $y$, whose detailed analytical
form is discussed in Ref. \cite{giacomo}.  Ref. \cite{giacomo}
discusses other walking prescriptions, and physical conditions
different from $2 < \mu < 3$ as well. For simplicity, in this paper we
discuss only the earlier walking prescription and the case where $\mu
< 3$, so as to create a strong departure from ordinary statistical
mechanics, namely from the condition where the second moment of
$\psi(\tau)$ is finite. We also set the condition $\mu > 2$ which
keeps the system far from the condition of a diverging first moment.
Why do we assign to $\psi(\tau)$ this condition of strong departure
from ordinary statistical mechanics?  We shall answer this important
question after making the reader familiar with the intriguing issue of
the memory emerging from the breakdown of the Poisson statistics. This
is a poorly understood property, in spite of the fact that 32 years
ago Bedeaux, Lindenberg and Shuler \cite{katja} wrote a clarifying
paper on this subject. We have seen that our method of analysis rests
on turning a time series into a diffusion process. If we imagine the
one-dimensional axis on which this diffusion process is realized as a
chain of infinite discrete sites, we can denote the state at time $ t
$ of the diffusing system through the vector ${\bf p}(t)$, with
$p_{i}(t)$ denoting the probability for the random walker to be at the
$i$-th site at time $t$. Thus, it is legitimate to ask the important
question of whether the knowledge of ${\bf p}(t)$ allows us to
determine ${\bf p}(t')$ with $t'> t$.  In other words, the question is
the following: does the information ${\bf p}(t)$ at a given time $t$
allow us to predict the state of the system at a later time? We want
to point out that the question refers to a set of random walkers, not
to a single random walker, whose walk, at a time scale larger than the
first moment of the waiting time distribution $\psi(\tau)$, is
certainly unpredictable.

A careful reading of Ref. \cite{katja} allows us to answer this
question with this apparently striking statement: this is possible
only in the Poisson case \cite{gerardo}. In the non-Poisson case an
infinitely extended memory emerges. The Poisson case is the only one
where the state ${\bf p}(t)$, with $t > 0$, determines the time
evolution of the system of interest. In all the other cases, the
future time evolution of the system also depends on ${\bf p}(t^{\prime
\prime})$, with $t^{\prime \prime}< t$. In other words, the system
time evolution retains memory of the initial condition ${\bf p}(0)$
forever.  It is evident that the concept of crucial event implies a
departure from the Poisson condition. In fact, the probability of
occurrence of a main-shock is expected to have memory, this
corresponds to the fact that the geophysical processes, responsible
for the main-shocks, do not generate random fluctuations, but
long-range correlation.

In the case of the terrorist network, we conjecture that the crucial
events, having either ideological or religious origin, and so
historical roots, are driven by non-Markovian master equations: This
implies that the irrelevant degrees of freedom, playing the role of a
thermal bath, are not equivalent to white noise, as it would be in the
case of Poisson statistics \cite{katja}. The choice of the condition
of $2< \mu < 3$ makes the resulting diffusion process depart
dramatically from the Gaussian state, thereby assigning to the crucial
events, either main-shocks or main secret events of the terrorist
network, a condition of striking departure from ordinary statistical
mechanics. In other words, we conjecture that the crucial events,
which, although invisible, influence cascades of secondary events, are
located in a basin of attraction of anomalous rather than normal
statistics, an assumption that fits the traditional wisdom of the
researchers in the field of complexity.  It has to be pointed out that
from a formal point of view a condition of infinite memory is realized
by $\mu < \infty$, without necessarily implying $\mu < 3$. However,
the condition $\mu > 3$ would not make the crucial event generate a
visibly anomalous diffusion, and an even more sensitive procedure
should be planned, to discover the existence of this kind of crucial
events. Thus, the condition $\mu < 3$, which, as earlier pointed out,
seems to be a plausible property of complex systems, corresponds to a
case where the procedure illustrated in this paper, is already
adequate, in the present form, to reveal their existence.  Before
ending this Section, we must clarify a problem that is a frequent
source of confusion. The infinite memory associated with non-Poisson
statistics might be mistaken as a way to make prediction.

We have to point out that the infinite memory is a concept referring
to probabilities, or to a set of walkers. The concept of Gibbs
ensemble, although fundamental for statistical mechanics, is based on
the assumption that many identical copies of the system are available
to us. Actually, we study only single systems. For instance, the
predictability of earthquakes, implies that knowing that a crucial
event occurred at time $t = t_{1}$, we can predict that the next will
take place at a time $t = t_{2} > t_{1}$. The time distance
$t_{2}-t_{1}$ cannot be predicted, if the events under study are
crucial in the sense earlier defined. However, if Eq. (1) were a
reliable model for the process under study, one might conjecture that
a specific observation of the geophysical motion is equivalent to
informing us about the new initial condition, after the back injection
taking place at $t = t_{1}$. The instant of the back injection is the
genuinely random event. The laminar motion ensuing this random event
is deterministic and, consequently, compatible with predictability, at
least in principle. Why do we leave room for randomness, in the moment
of selection of the new initial condition? This is equivalent to
associating crucial events to randomness, and a thorough discussion of
this issue beyond the limits of this paper. If we adopt the usual view
that randomness is an expression of our ignorance about the infinitely
many and irrelevant degrees of freedom in a system, this choice is
equivalent to a drastic simplification of the problem under
study. Going beyond that would be equivalent to predict the occurrence
time of main-shocks, in the case of seismic fluctuations, and of
terrorist actions, in the case of the war to terrorism. For the time
being, our purpose is much less ambitious.

\section{memory beyond memory}

Notice that the title of the paper of Ref. \cite{memorybeyondmemory},
\emph{memory beyond memory}, is probably incomprehensible to all those
who do not know the fundamental work of Ref. \cite{katja}. On the
basis of the results of Ref. \cite{katja} we can explain the meaning
of this title. The majority of events under observation in
Ref. \cite{memorybeyondmemory} are not crucial events. The crucial
events, which are rare, are imbedded in a sea of secondary events,
also called pseudo-events. These secondary events are influenced by
the crucial events and play a camouflage role that makes the really
crucial events invisible. As a consequence of being secondary, the
pseudo-events have memory of the crucial events influencing them.

It is worth remarking that in the case of seismic events the crucial
events are the main shocks and the secondary events are the Omori
swarms of aftershocks.  We make an important conjecture: the case of
the terrorist network rests on the picture of the passive supporters
of terrorist activities.  These supporters generate events that,
although secondary, are dependent on the main terrorist events, of
which they bear memory. This is the memory of the second kind, the
memory of first kind being, as pointed out in the earlier Section, the
memory corresponding to the non-Poisson statistics of the crucial
events. The terrorists trigger events, with either religious or
ideological memory, and these crucial events influence secondary
events, the action of passive supporters, which are characterized by
\emph{memory beyond memory}. Although, the term \emph{memory beyond
memory} was originally coined for the completely different purpose of
helping the search for the key physiological processes behind
heart-beating \cite{memorybeyondmemory}, we find it to be especially
adequate to describe a procedure of statistical analysis aiming at
helping the Intelligence Community in the war against terrorism.  For
all these reasons, it is convenient to review a model that was
originally proposed to illustrate the origin of memory of the second
type \cite{memorybeyondmemory}. The model consists of two particles.
The first particle is the visible one. We maintain the same dynamic
rule as that established by Eq. (\ref{dynamics}), though now we change
the role of this particle from the generator of crucial events to the
generator of pseudo-events. This means that we keep using the visible
particle to generate events, and consequently the time series to
study, with a different back injection rule, though, for the purpose
of generating events that are not random.  To do that, following
Ref. \cite{memorybeyondmemory}, we introduce a second particle ruled
by an equation of the same kind as Eq. (\ref{dynamics}). However,
while Eq. (\ref{dynamics}) refers to events that we can monitor, now
the second particle refers to hidden events. Therefore we refer to
this particle as the invisible particle. In conclusion, we describe
this model by means of the following set of equations:
\begin{eqnarray}
\label{dynamics2}
\frac{dx_{vis}}{d t} = \alpha x_{vis} ^{z}~~~,
\\
\frac{dx_{invis}}{d t} = \beta x_{invis}^{\zeta}~~~.
\end{eqnarray}
We assume that the dynamics of the visible particle are much
faster than the dynamics of the invisible particle. Thus, the visible
particle gets to the border and  is injected back many times before the
occurrence of the leading, or crucial, event. The crucial event occurs
when the invisible particle reaches the border and  is injected back
randomly to a new initial condition in the interval $I$. Before the
occurrence of this crucial event, the visible particle has been
injected back following a very simple deterministic prescription. In
the earlier work of Ref. \cite{memorybeyondmemory} to check the
efficiency of our method of analysis we have made the assumption that
the initial condition is always the same, and it is changed randomly
only when the invisible particle is injected back.

At this stage, we wonder if it is possible to distinguish the crucial
events from the surrounding pseudo events; in particular we wonder if
a statistical method of analysis exists that detects the waiting time
distribution of the crucial events.  The answer is positive, and can
be found in the paper of Ref.\cite{memorybeyondmemory}. First of all,
we have to convert the time series into a diffusion process. According
to the prescriptions of Ref. \cite{giacomo}, we evaluate the Shannon
entropy of this diffusion process. This is why this technique of
analysis is called Diffusion Entropy (DE) method. As is well known,
the distance $x$ travelled by the walkers is related to time by the
relation $x \propto t^{\delta}$, where $\delta$ is termed diffusion
index. If the diffusion process is not the sum of uncorrelated
fluctuations, the scaling parameter $\delta$ departs from the
prescription of ordinary statistical mechanics, namely, from $\delta =
0.5$.  The DE method is an efficient way to determine $\delta$
\cite{giacomo}.

  In the case where the visible events are not correlated, and the
walking rule of Section II is adopted, the scaling $\delta$,
determined by means of the DE method, and the power index $\mu$ of the
waiting time distribution $\psi(\tau)$ of Eq. (5), are related by
means of Eq. (9).  The violation of this crucial condition suggests
that the events under observation are not genuine events, but rather
pseudo-events, bearing, as a consequence of that, memory of the second
type. Actually, the way to proceed is as follows. We evaluate
numerically the waiting time distribution $\psi_{exp}(\tau)$, by
running the two-walkers model. The observation of visible events
determines the waiting time distribution
\begin{equation}
\label{waitingtimedistributionprime}
\psi_{exp}(\tau) = (\mu^{\prime} -1) \frac{T^{\mu^{\prime}-1}}{(T + \tau)^{\mu^{\prime}}}.
\end{equation}
We do not address here the interesting problem of establishing $\mu'$
as a function of the parameter of the two-walkers model.  This is not
crucial for the discussion of this paper.  Let us limit ourselves to
noticing that $\mu' > \mu$.  The numerical results of
Ref. \cite{memorybeyondmemory} show that $\delta$ does not have
anything to do with $1/(\mu^{\prime}-1)$.  These numerical results
rather prove the attractive fact that Eq. (\ref{scaling}) applies, but
with $\mu$ denoting the power coefficient of $\psi(\tau^{[m]})$, and
$\tau^{[m]}$ the time distance between two consecutive crucial events
(the subfix $m$ here stands for main events, in analogy with the
definition used in \cite{earlier}).  In other words, the scaling
coefficient $\delta$, detected by means of the DE method, reveals an
important statistical property of crucial and invisible events.  Let
us summarize the procedure that we propose to detect the statistical
properties of invisible and crucial events.  First of all, we adopt an
experimental view, and we derive from the real sequence under study
the waiting time distribution $\psi_{exp}(\tau)$, referring to the
time distance between two consecutive events.  If we find that the
waiting time distribution is not exponential, we have a first
indication of complexity.  If the distribution is an inverse power
law, we record the power law index, $\mu^{\prime}$.  Then we use the
DE method to measure the scaling parameter $\delta$.  The condition
$\delta = 1/(\mu^{\prime}-1)$ is a plausible indication that we are
observing a time sequence of significant events.  If $\delta$
significantly departs from $1/(\mu^{\prime}-1)$, there are good
reasons to believe that
\begin{equation}
\label{truescaling}
\mu = 1 + \frac{1}{\delta} 
\end{equation}
is a reliable indicator of the complexity of invisible and crucial
events.  It is important to note that this important conclusion is
supported by the analytical treatment of Ref. \cite{memory2}. The
authors of Ref. \cite{memory2} shows that in the long-time limit the
memory of the pseudo-events is lost, and the process under study
becomes equivalent to a L\'{e}vy flight, corresponding to the power
index $\mu$ of the crucial events.

\section{The Omori's law as a source of pseudo-events}

\begin{figure}[!h]
\includegraphics[width=10.5 cm]{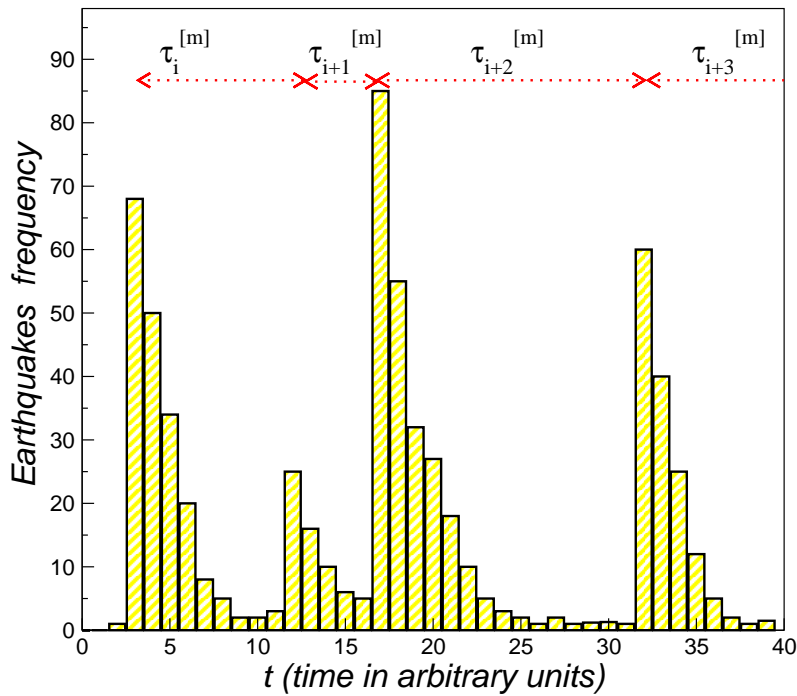}   
\caption{\label{fig1} We report a schematic figure illustrating the
typical earthquakes frequency vs time.  In correspondence to each
main-shock we observe a frequency peak determined by the after-shock
swarm.  The peaks decay according to the Omori's law, see text.  The
horizontal dotted arrows indicate the time intervals $\tau_i^{[m]}$
between two consecutive main-shocks. The DE method gives information
on the distribution of these time intervals.}
\end{figure}

In this and in the next Section we review the work of
Ref. \cite{earlier} for the main purpose of proving that the results
of this paper are a realization of the method for the search of
invisible and crucial eventes illustrated in Section III.  In
Fig. \ref{fig1} we report the sketch of the typical earthquakes
frequency \emph {vs} time in the catalog that we shall consider in the
next Section. By $\tau_{i}=t_{i+1} - t_{i}$ we indicate the time
interval between an earthquake and the next.  Each peak of frequency
(cluster) in figure includes the time location of a main-shock.  The
time interval between one peak and the next is reported in figure and
is denoted by the symbol $\tau_{i}^{[m]}$, where the superscript $m$
stands for main-shock, since the main-shocks are the main event in the
case of seismic fluctuations.  According to the definition of crucial
events given in Section I, we must make the assumption that two
different $\tau^{[m]}$'s are not correlated, i.e.  that the
correlation function is:
\begin{equation}
\label{crucial}
\langle \tau_{i}^{[m]} \tau_{j}^{[m]} \rangle = \left \langle
  \left ( \tau^{[m]} \right )^2 \right \rangle \delta_{i,j}.
\end{equation}
Note that with the symbol $\tau^{[m]}$ we denotes distances between
two consecutive crucial events. Thus the corresponding waiting time
distribution is equivalent to that distribution of Eq. (5).

The experimental determination of this distribution would imply the
adoption of a way to identify the main-shocks. Although the geologists
might suggest reliable criteria for their identification, for instance
through the magnitude, with the use of our method we can determine
their statistical properties without identifying them. Thus, the
symbols $\tau^{[m]}$ denote distances between consecutive events that
are assumed to be invisible.  One of the models adopted to describe
the time distribution of earthquakes is the Generalized Poisson (GP)
model \cite{shlien,gardner,gasperini,console,godano}.  Basically the
GP model assumes that the earthquakes are grouped into temporal
clusters of events and these {\it clusters are not long-range
correlated}: in fact the clusters are distributed at random in time
and therefore the time intervals between one cluster and the next one
follow a Poisson distribution.  On the other hand, the {\it
intra-cluster earthquakes are correlated} in time as it is expressed
by the Omori's law \cite{omori,utsu}, an empirical law stating that
the main-shock, i.e. the highest magnitude earthquake of the cluster,
occurring at time $t_{0}$ is followed by a swarm of correlated
earthquakes (after-shocks) whose number (or frequency) $n(t)$ decays
in time as a power law, $n(t) \propto (t-t_{0})^{-p}$, with the
exponent $p$ being very close to $1$.  The Omori's law implies
\cite{bak} that the distribution of the time intervals between one
earthquake and the next, denoted by $\tau$, is a power law $\psi(\tau)
\propto \tau^{-p}$.  This property has been recently studied by the
authors of Ref. \cite{bak} by means of a unified scaling law for
$\psi_{L,M}(\tau)$, the probability of having a time interval $\tau$
between two seismic events with a magnitude larger than $M$ and
occurring within a spatial distance $L$.  This has the effect of
taking into account also space and extending the correlation within a
finite time range $\tau^{*}$, beyond which the authors of
Ref. \cite{bak} recover Poisson statistics.  Let us discuss the GP
model in the light of the general remarks of Sections II and III. The
Poisson assumption about the distribution of time distances between
one main-shock and the next is equivalent to assigning no memory to
the geophysical process responsible for the main-shock. This conflicts
with our definition of crucial events and with our conviction that the
crucial events, as unpredictable as the time duration of a laminar
region is, cannot be determined by erratic bath fluctuations. The GP
model, if supported by the statistical analysis of data, would imply
that our definition is not correct, and that crucial events can be
generated also from within ordinary statistical mechanics. This would
conflict also with the tenets of complexity, which seem to connect
cooperation and inverse power law relaxation. In Section V we shall
prove that the GP model must be dismissed. In fact, using the method
of statistical analysis reviewed in this paper, it is shown
\cite{earlier} that the asymptotic scaling generated by the GP scaling
would be $\delta = 0.5$, which in fact corresponds to the prescription
of Section II, when $\mu > 3$. It is worth recalling that the Poisson
condition sets the exponential decay of $\psi(\tau)$, and thus $\mu =
\infty$. The statistical analysis of real data, discussed in Section
V, will prove that $\delta = 0.94$, that the GP model is incorrect,
and that we cannot rule out the possibility that the main-shocks
fulfill our definition of crucial events.
 
\section{Data and results}

The catalog we have studied covers the period 1976-2002 in the region
of Southern California spanning $20^0$ N -$45^0$ N latitude and
100$^{0}$ W 125$^{0}$ W longitude \cite{scsn}.  This region is crossed
by the most seismogenetic part of the San Andrea fault, which
accommodates by displacement the primarily strike-slip motion between
the North America and the Pacific plates, producing velocities up to
$47 mm/yr$ \cite{turcotte}.  The total number of recorded earthquakes
in the catalog is $383687$ and includes the June 28 1992 Landers
earthquakes (M = 7.3), the January 17 1994 Northridge earhquake (M =
6.7), and the October 16 1999 Hector Mine earthquake (M = 7.1).
Geophysical observations point out that these large earthquakes have
triggered a widespread increase of seismic events at remote distances
in space and in time \cite{hill,parsons}.  The coupling of the sources
of stress change (i.e. large earthquakes occurrence) and seismicity
triggering mechanisms is a primary target of geophysical
investigations, and, as shown below, is revealed by the DE analysis.
In Fig. \ref{fig2} we report the results of the DE method.  The
analysis was performed by setting $\xi(t) = 1$ when an earthquake
occurs at time $t$ (independently of whether it is a main-shock or an
after-shock), and $\xi(t)=0$ if no earthquake happens.  By means of
the full circles we denote the entropy $S(t)$ as a function of time
when all the seismic events of the catalog are considered
(independently of their magnitude $M$).  After a short transient, the
function $S(t)$ is characterized by a linear dependence on $\ln t$.  A
fit in the linear region gives a value of the scaling parameter
$\delta = 0.94 \pm 0.01$ at $95 \%$ of confidence level.  We next
consider only the earthquakes with magnitude larger than a fixed value
$\bar M = 2, 3, 4$.  We see that, regardless of the value of the
threshold $\bar M$ adopted, the function $S(t)$ is characterized by
the same long-time behavior with the same slope.  This indicates that
we are observing a property of the time location of large earthquakes.
This leads us to conclude that the time intervals between two large
events fit the distribution of Eq.(5), with the value of $\mu$ related
to $\delta$ through Eq.(9), $\mu = 2.06 \pm 0.01$. Our conclusion is
also supported by other two numerical analyses based on different
prescriptions to construct the diffusion process.  The former rests on
assuming $\xi(t)$ equal to the magnitude $M$ of the earthquake, at
each time when an earthquake occurs.  The latter sets with equal
probability either $\xi(t) = 1$ or $\xi(t) = -1$ when an earthquake
occurs \cite{giacomo}.  Both the methods give the same exponent
$\mu=2.06 \pm 0.01$.  In conclusion, the statistical analysis of real
data rules out the GP model, which would conflict with our definition
of crucial events.  Thus, there is still room for the main-shocks to
fit our definition of crucial events. The authors of
Ref. \cite{earlier} prove that, under the stationary condition, they
do.  Are these crucial events also invisible? This is a question of
fundamental importance for the war against terrorism. The answer is
that the main-shocks are not quite invisible. For professional
geologists it is possible to identify all of them. However, our method
of analysis works, regardless of whether the crucial events are
invisible or not. In fact, Fig. 2 (a) shows that the asymptotic
properties of the entropy indicator are independent of the threshold
$M$ adopted. Since the magnitude is a distinctive property of the
main-shocks, we conclude that the results of our analysis do not imply
that the main-shocks are identified. This is the reason of our
conviction that our method can be successfully used even when the
crucial events are quite invisible.

\begin{figure}[!h]
\includegraphics[width=10.5 cm]{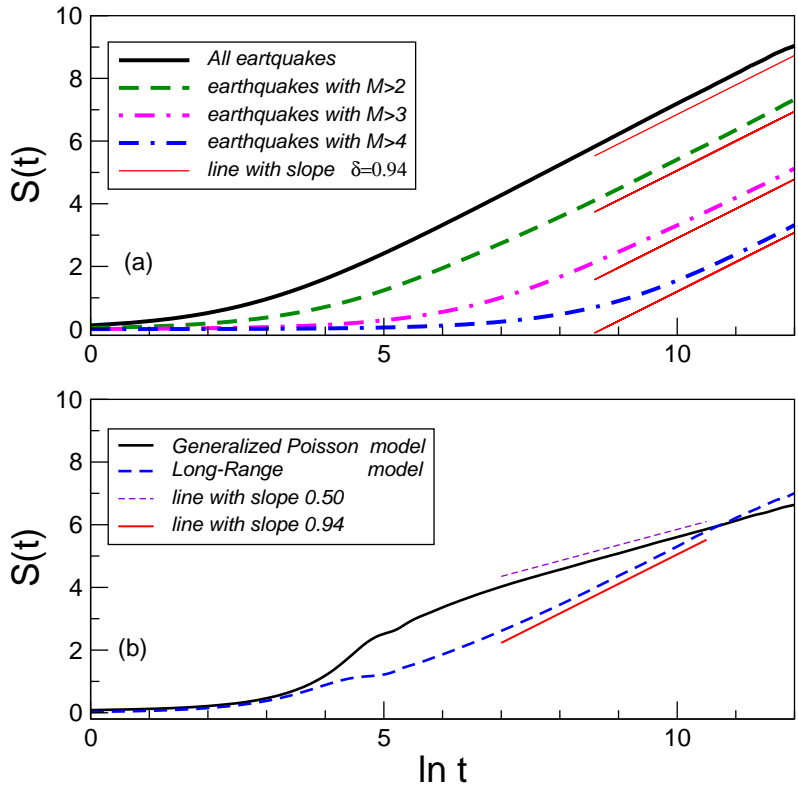}   
\caption{\label{fig2} 
(a) We plot the Shannon entropy $S(t)$ of the diffusion process as a function of
time (in minutes), in a logarithmic time scale. From top to bottom, the curves 
refer to all seismic events without considering the magnitude $M$ and to events with 
magnitude greater than $\bar M = 2,3,4$ respectively. 
The straight lines are plotted to guide the eye and have the slope $\delta = 0.94$.
(b) We plot the Shannon entropy S(t) of the diffusion process as a function
of time, in a logarithmic time scale for the GP and of the LR model.   
We plot also  two straight lines with  slopes $\delta =
0.5$ and $\delta = 0.94$, see text for further details.}
\end{figure}

\section{Generalized Poisson (GP) and Long-range (LR) model}

We now illustrate how the DE method works on two artificial
earthquakes time series: the first generated by means of the GP model,
and the second generated by a new model, the Long-Range (LR) model,
that we propose as a better model to reproduce the properties of the
catolog considered.  In the LR model the earthquakes are grouped into
temporal clusters, and, as in the GP model, the number of earthquakes
in a cluster follows the Pareto law, i.e. a power law distribution
with exponent equal to 2.5 \cite{console}. The events within the same
cluster are distributed according to the Omori's law: the interval
$\tau$ follows a power law with exponent $p=1$.  However, in the LR
model the time distance $\tau^{[m]}$ between one cluster and the next
follows a power law with exponent $\mu=2.06$, rather than a Poisson
prescription as in the GP.  Notice that this value of $\mu$ is close
to the border between stationary and non-stationary condition
\cite{giacomo}.  The two sequences have the same time length.  We
choose the number of clusters in order to have the same total number
of earthquakes as in the real data \cite{note1}.  The result of the DE
on the artificial sequences is reported in Fig. 2(b).  The GP model is
characterized by a long-time behavior that, as expected, fits very
well the prescription of ordinary statistical mechanics, with $\delta
= 0.5$. The LR model yields the quite different scaling $\delta =
0.94$. It is also clear that the LR model yields a behavior
qualitatively similar to that produced by the real data of Fig. 2 (a)
as well as the same scaling parameter $\delta = 0.94$, while the GP
fail reproducing both properties.  Note that from the reasons why the
DE method reveals the genuine statistical properties of the crucial
events in this case, are different from those justifying the
efficiency of the method in the case of the model of Section III. In
this case, the true reason seems to be that for large distances
between one main-shock and next the pseudo-events tend to concentrate
immediately after the last main-shock with a so low density
immediately before the occurrence of the next as to create a condition
where the number of pseudo-events is essentially independent of the
length of the laminar region. In the case of the model of Section III,
on the contrary, the number of pseudo-events is proportional to the
length of the laminar region.
\section{Conclusions}

It is the time for us to balance the detection of crucial and
invisible events in general, with possible applications to the war
against terrorism. Our definition of crucial events implies a strong
departure from Poisson statistics.  In the recent literature there is
a general agreement about the fact that complex networks, including
the terrorist network, are scale-free systems. The authors of another
paper of these Proceedings \cite{linguistics} show that there exists a
connection between the scale-free condition and non-Poisson
statistics. This seems to support the definition of crucial event
adopted in this paper. On the other hand, there exists an interesting
connection with the Small Words theory illustrated by Latora and
Marchiori \cite{vito}, a paper of these Proceedings, explicitly
devoted to the war against terrorism. This paper, in turn, rests on a
perspective that is related to the sociological picture illustrated
Elliott and Kiel \cite{euel} for the same purpose. It is worth
mentioning that a local version of the DE method can be applied to the
timely detection of toxicants \cite{massi}.  In conclusion, the
present paper belongs to a set of contributions to these Proceedings
\cite{linguistics,vito,euel,massi}, which might bear benefits to a
program of research to combat terrorism.  As to the detection of
invisible and crucial events for specific purpose of the war to
terrorism, the procedure to follow depends on the data to analyze, and
on the forms, under which they will be made available to the
investigators. Nevertheless, with the present paper, we are convinced
that at least the first few steps of the search for crucial and
invisible events, take a clear shape. It seems to be evident what the
first step of this procedure will be the evaluation of
$\psi_{exp}(\tau)$, and of the corresponding power index, denoted by
the symbol $\mu^{\prime}$ in this paper. The second step will be the
evaluation of $\delta$, by means of the DE method, and the comparison
of $\delta$ with $1/ (\mu^{\prime}-1)$.  If the two values do not
coincide, and the difference is larger than the statistical error,
this has to be thought of as a plausible indication that there are
invisible and crucial events involved. Then, we shall have to decide
whether or not recourse can be done to the simple prescription of
Eq.(13) to establish the degree of complexity of the invisible crucial
events. This will require further research work determined by the
specific nature of the data to analyze.

PG acknowledges support from ARO, through Grant DAAD19-02-0037.


\begin{thebibliography}{99}
\bibitem{earlier} M. S. Mega, P. Allegrini, P. Grigolini, 
V. Latora, L. Palatella, A. Rapisarda, and S. Vinciguerra, Phys. Rev. Lett. 90, 188501 (1-4) (2003). 
\bibitem{memorybeyondmemory} P. Allegrini, , P. Grigolini, P. Hamilton, L. Palatella, and G. Raffaelli,  Phys. Rev. 65, 041926-1-5 (2002).
\bibitem{memory2} P. Allegrini, R, Balocchi, S. Chillemi, P. Grigolini, P. Hamilton, R. Maestri, L. Palatella, amd G. Raffaelli,  Phys. Rev. E 67, 062901 (2003). 
\bibitem{giacomo}
 P. Grigolini, L. Palatella, G.
Raffaelli, Fractals, 9, 439-449 (2001).
\bibitem{katja} D. Bedeaux, K. Lakatos Lindenberg, and K. E. Shuler,
J. Math. Phys. 12, 2116 (1971).
\bibitem{gerardo} P. Allegrini, G. Aquino, P. Grigolini, L. Palatella,
A. Rosa, submitted to Phys. Rev. E.
\bibitem{shlien} S. Shlien, M.N. Toksoz, 
Bull. Seism. Soc. Am., 60, 1765 (1970).
\bibitem{gardner} J.K. Gardner and L. Knopoff, 
Bull. Seism. Soc. Am., 64, 1363 (1974). 
\bibitem{gasperini} P. Gasperini, F. Mulargia, 
Bull. Seism. Soc. Am., 79, 973 (1989).
\bibitem{console}
R. Console, M. Murru, J. Geophys. Res., 106, 8699, (2001).
\bibitem{godano}
C. Godano and V. Caruso, Geophys. J. Int. 121, 385 (1995).
\bibitem{omori} 
F. Omori, J. College Sci. Imper. Univ. Tokyo  7, 111 (1895).
\bibitem{utsu}  
T. Utsu, Geophys. Mag., 30, 521 (1961).
\bibitem{bak} P. Bak, K. Christensen, L. Danon and T. Scanlon, 
 Phys. Rev. Lett {88}, 178501 (2002).
\bibitem{scsn} 
The catalog has been downloaded from 
the Southern California Earthquake Data Centre 
http://www.scecdc.scec.org/ftp/catalogs/SCSN/ 
\bibitem{turcotte} 
D. L. Turcotte and G. Schubert, 
Geodynamics, Cambridge University Press 2002. 
\bibitem{hill} D.P. Hill et al., Science 260, 1617 (1993).
\bibitem{parsons} T. Parsons, J. Geophys. Res., 107, 2199 (2001).
\bibitem{note1} 
We made other tests by adopting different exponents of 
the Pareto's law ($3.0 \pm 0.7$) and different 
number of clusters, obtaining different transient behavior but 
the same value of $\delta$.
\bibitem{linguistics} P. Allegrini, P. Grigolini, L. Palatella, Chaos, Solitons and Fractal, Proceedings of the October 2002 Denton Workshop
\bibitem{vito} V. Latora. M. Marchiori, "How the Science of Complex Networks Can Help 
Developing Strategies Against Terrorism", Chaos, Solitons and Fractal, Proceedings of the October 2002 Denton Workshop.

\bibitem{euel}E. Elliott, D. Kiel,  "A Complex Systems Approach for Developing Public Policy Toward Terrorism: An Agent-Based Approach", Chaos, Solitons and Fractal, Proceedings of the October 2002 Denton Workshop.

\bibitem{massi} M. Ignaccolo, P. Grigolini, G. Gross, "Towards the timely detection of toxicants", Chaos, Solitons and Fractal, Proceedings of the October 2002 Denton Workshop.
\end{thebibliography}
\end{document}